\documentclass[twocolumn,prb,showpacs,preprintnumbers,amsmath,amssymb]{revtex4}
\usepackage{graphicx}
\usepackage{dcolumn}
\input epsf

\begin{document}

\title{Electronic correlations and disorder in transport through 
one-dimensional nanoparticle arrays}
\author{E. Bascones$^{1,2,3,*}$, V. Est\'evez$^{1}$, J.A. Trinidad$^2$ and 
A.H. MacDonald$^2$}
\affiliation{$^1$Instituto de Ciencia de Materiales de Madrid, CSIC,
Cantoblanco, E-28049 Madrid, Spain
 \\$^2$ Department of Physics, University of Texas at Austin, Austin 
TX-78712, USA \\ $^3$Theoretische Physik, 
ETH-H\"onggerberg, CH-8050 Zurich, Switzerland}
\date{\today}
\begin{abstract}
We analyze and clarify the  transport properties of a one-dimensional 
metallic nanoparticle 
array with interaction between charges  restricted to  charges 
placed in 
the same conductor. 
We study the threshold voltage, 
the I-V curves  and the potential drop through the array and their dependence 
on the array parameters including the effect of charge and resistance disorder.
We show that very close to threshold the current depends linearly on voltage 
with a slope independent on the array size. At intermediate bias voltages, for 
which a Coulomb staircase is observed we find that the average potential drop 
through the array oscillates with position. At higher voltages I-V curves are 
linear but have  a finite offset voltage. We show that the slope is given by 
the inverse of the resistances added in series and estimate the voltage at 
which this linear regime is reached. We also calculate the offset voltage and 
relate it to the potential drop through the array.
\end{abstract}
\email{leni@icmm.csic.es}
\maketitle

Nanoparticle arrays made of metallic\cite{metallicwhetten,metallicheath,metallicjaegerdis,metallickiehl,metalliclin,metallicjaegerquasi1d,metalliczhang,metallicjaegernature,metallicschoenen,metallicysemi}, semiconducting\cite{metallicysemi,semimurray,semitalapin,semibawendi,semisionnest,semidrndic,semiheath}, magnetic\cite{magneticsun,magneticblack,magneticpuntes} or combined\cite{mixedredl,mixedkorgel,mixedshevchenko} materials and
with radii of the order of 2-7 nm can be now synthesized. The 
transport 
properties of these systems are influenced by the ratio between the energy 
level spacing, the charging energy of the nanoparticles, and the temperature. 
The first two
quantities depend on the material and the size of the nanoparticle. In the 
case of metallic nanoparticles, at not too low temperatures, the level spacing 
is much smaller than the temperature and does not play any role in the 
transport\cite{singlecharge}.
On the contrary, the charging energy is of the order of 0.1 eV. Strong 
interactions between the electric charges and the possibility of tuning 
interparticle coupling make nanoparticles arrays an ideal 
system 
to study correlated 
motion\cite{collectiveheath1,collectiveheath2,collectiveheathpb,collectivebard,collectivekorgel,collectivepileni,collectivesheng,Middleton93,Matsuoka98,Shin98,Kaplan02,Kaplan03,Kinkhabwala04,DasSarma,Belobodorov,Glazman,Turlakov,Efetov,Shklovskii}.

Experimentally, these arrays are strongly influenced by 
disorder\cite{disorderjaeger,disorderheath,disordercordan}. Local 
charging disorder is present in all arrays due to randomly dispersed charged 
impurities lodged in the substrate or in the materials that separate and 
surround the 
nanoparticles. Because of the exponential dependence of the tunneling 
resistance, 
even a small dispersion in the 
distance between nanoparticles results in large variations in the 
tunneling resistances of the junctions. Differences in the island sizes and 
voids in the lattice can be other sources of disorder\cite{metallicjaegerdis}.

Due to the combination of disorder and charging effects the current in voltage 
biased arrays is blocked up to a threshold 
voltage\cite{disorderjaeger,Tinkham99,Ancona01,Bakhalov89,Middleton93,Hu94,Melsen97,Berven01,Kaplan03,Schoeller05} $V_T$. For bias voltages larger than $V_T$ current is 
in general non-linear in voltage with a power-law dependence\cite{disorderjaeger, Tinkham99,Clarke98} close to threshold, a 
linear dependence recovered at high-voltages and frequently a step-like behavior,  called a Coulomb staircase, at 
intermediate voltages. 
Most studies have focused on the statistical analysis of the 
threshold voltage
and on the power-law behavior of the current close to this 
threshold. 
This exponent depends on the dimensionality of the array, but there is 
controversy 
between different theoretical approaches in the one-dimensional case, with
both linear\cite{Middleton93,Kaplan03,Jha05} and 
square-root\cite{Reichhardt03} predictions.
Much experimental work has been concentrated on two and three dimensional 
arrays, but some quasi-one dimensional systems have also been 
fabricated\cite{metallickiehl,metallicjaegerquasi1d,Tinkham99}.
Comparison between experiments and theory is not yet well settled.

In this paper we provide a complete description of the zero-temperature 
transport properties
of one-dimensional metallic
nanoparticle arrays for the case in which interactions are 
restricted to charges in the same nanoparticle (onsite limit).
In a separate work\cite{paperlongrange} we discuss the effect of the 
long-range character of the interactions on these properties. 
We resolve the controversy on the power-law, analyze carefully dependencies on
 different array parameters and  estimate the voltage which delimits each 
of the transport regimes to favor comparison with experiment.
We discuss arrays with and without charge disorder. Although clean arrays 
are mainly of academic interest, their analysis will help us to understand the 
main features of the experimentally more relevant, disordered arrays. The 
effect of variations in the junction resistances is also analyzed.  
Disorder in capacitances (nanoparticle size variations) is not considered as it
is less important in present  experiments. 
Nanoparticles 
synthesized nowadays are monodispersed in size to a few percent. 
In any case, the effect of capacitance disorder\cite{cap} in most
of the properties studied here
can be deduced from the analytic approximations provided in the text.   
Due to the one-dimensionality of the
array and the nearest neighbor tunneling considered we assume that there are
no nanoparticle voids in the array, as this would completely prevent current 
flow. 
We have analyzed the threshold voltage, the I-V characteristics, close to 
threshold but also at larger bias voltages and the potential drop through the 
array.
Among the results presented we show that the threshold voltage of clean arrays
in the onsite case does not approach the one found for capacitively coupled 
junctions in the limit of weak coupling\cite{Bakhalov89,Hu94}.
The controversy in the power-law dependence of current on bias voltage close 
to threshold is clarified and shown to be linear, but with a slope different 
from the one previously predicted\cite{Jha05}. 
We find signatures of correlated charging of the system in the potential drop 
through the array, especially  at intermediate values of the bias voltage. 
We show that even at high voltages, with the conductance controlled by the sum 
of the resistances in series, the potential drop through the array is not 
completely linear, but it has a contribution due to charging effects. 
We identify an asymmetry external parameter $\alpha$ which controls the bias 
voltage drop. The influence of $\alpha$ has barely been
discussed in previous 
works. We show that its effect can {\it a priori} be observed experimentally in the width of the Coulomb staircase steps. 
Calculations are perfomed numerically, but analytic approximations are given in
several limits and compared with numerical results. 

\section{THE MODEL}
\label{model}
We analyze the current through a one-dimensional array of $N$ metallic 
nanoparticles placed between two electrodes. We consider the classical 
Coulomb blockade regime with $\delta \ll K_BT <E^{isl}_c$. $\delta$ is the 
level 
spacing, $T$ the temperature and $E^{isl}_c=1/(2C^{isl})$ the charging 
energy of the 
islands, with capacitance $C^{isl}$. Here and in the following the electronic charge $e=1$
We assume that each nanoparticle has a continuum level spectrum ($\delta=0$) 
and a constant density of states at the Fermi level, but a gap $E^{isl}_c$ 
for 
adding charge. 
We restrict electrostatic interactions to 
those charges on the same conductor: capacitive coupling vanishes. This limit is referred to as short-range 
or onsite interaction limit.
 The nanoparticles are separated by high tunneling barriers with a resistance 
much larger than the quantum of resistance. In these conditions the charge in 
the islands can be assumed fixed and quantized. Eventually we allow tunneling 
processes between nearest neighbors, and treat the transport at the sequential 
tunneling level. A single charge is involved in the tunneling process.
We assume that when a 
charge hops, the charge density in the final state of the array immediately relaxes to 
the electrostatic equilibrium configuration.    

 We take into account that the electrodes are not ideal voltage sources, but 
have a finite self-capacitance. In equilibrium, and before the tunneling event 
the electrodes are held at a given potential due to the charge provided by a 
battery. We assume that the tunneling time, i.e. the time needed by the 
electron to cross the tunnel barrier, is smaller than the circuit 
characteristic time that determines how quickly the battery can transfer 
charge to the leads in order to restore the voltage at the electrodes. 
As a consequence, 
just after the tunneling process the electrodes will not necessarily
be at the same 
potential at which they were at before the tunneling event because the charge, 
provided by the battery, necessary to restore their initial potentials has not
arrived yet. 
 The voltage is restored to the nominal 
value before the next tunneling event.   
For finite-range interactions the potentials on the leads will 
thus fluctuate in response to 
all tunneling events, even those that do not directly involve the electrodes. 
In the short-range case, considered here, they will fluctuate only when an 
electron jumps into or out of the leads.

The current is calculated numerically by means of a Monte Carlo simulation,
described in the Appendix, which depends on the tunneling rates.
The probability of a tunneling process\cite{singlecharge} is given by
\begin{equation}
\Gamma{(\Delta E)}=\frac{1}{R}\frac{\Delta E}{exp(\Delta E/K_B T)-1}
\label{tunnrate}
\end{equation}
with $R$ the tunneling resistance of the junction. We will restrict the
discussion to zero temperature for which $\Gamma(\Delta E)=-\Delta E/R \Theta(-\Delta E)$.
$\Delta E$ is the difference between the energy of the system before and
after the tunneling event, with the sign convention that $\Delta E$ is negative if the energy decreases. 
It excludes
the work done by the battery to recharge the electrodes, as explained before. The energy gained by
 tunneling is assumed to be dissipated. Only changes in energy with
electrostatic origin are considered. The energy of our system is given by
\begin{equation}
F=\frac{1}{2}\sum_{\alpha=0}^{N+1}\frac{Q^{2}_\alpha}{C_\alpha} +
\sum_{i=1}^{N}Q_i \phi_i^{dis}
\end{equation}
Labels $0$ and $N+1$ refer to source and drain electrodes and $1$,...,$N$
to the islands. In the following, latin capital and lower case letters
are used to denote electrodes and islands respectively. Greek indexes will
be used when the labels refer to both islands and electrodes. $C_\alpha$ and 
$Q_\alpha$ are respectively the conductor capacitance and the charge 
in it.
The charges provided by the battery at the source and drain electrodes, which 
maintain their potentials at $V_0$ and $V_{N+1}$ are $Q_0=C_0 V_0$ and 
$Q_{N+1}=C_{N+1} V_{N+1}$. The capacitances of the electrodes are much larger
than those of the nanoparticles.
$\phi_i^{dis}$
is a random potential at each island due to randomly dispersed charges within 
the substrate and within the material surrounding the 
nanoparticles\cite{Xue03}. 
Clean arrays will be characterized by $\phi_i^{dis}=0$ for every $i$.
The random potential is included only at the islands because a similar term at 
each electrode is compensated by the battery and thus has no effect on 
transport. In the case of disordered arrays, the disorder potentials can, in
principle, take values larger than the charging energy $E^{isl}_c$. 
However, for
large values of the disorder potential, charges flow to compensate for these 
large fluctuations. In the case of short-range interactions, except if the
original disorder potential is very weak, once the screening
of the potential due to the mobile charges is taken into account, the set of 
disorder potentials is uniformly distributed in the interval 
$-E^{isl}_c \leq \phi_i^{dis} \leq E^{isl}_c$\cite{disorderjaeger}, and 
in the following we consider this distribution. 

The relevant quantity for the transport is the change in energy due to a 
tunneling event.
The tunneling process can be seen as the creation of a hole in 
the conductor $\alpha$ from which the charge leaves, 
$Q_\alpha \rightarrow Q_{\alpha}-1$, and the addition of an electron in 
$\beta$ at which the charge arrives, $Q_{\beta} \rightarrow Q_{\beta}+1$. Here 
and thereafter, we let $+1(-1)$ denote the charge of an electron (hole). 
In fact, the change in energy 
 can be rewritten as the energy to create an electron-hole 
(also called in the 
following excitonic energy) plus the difference in potential 
between the sites 
involved in the process before the tunneling event.
\begin{equation}
\Delta E = E^{e-h}_{\alpha,\beta} + (\phi_\beta -\phi_\alpha)
\label{change} 
\end{equation}
The first term gives the energy to create an electron-hole pair in an 
uncharged clean array and is given by
\begin{equation}
E^{e-h}_{\alpha,\beta}= \frac{1}{2 C_\alpha}+
\frac{1}{2 C_\beta}=E_c^{\alpha}+E_c^\beta
\label{excitonic}
\end{equation}
This energy does not depend on the direction of tunneling (from $\alpha$ to 
$\beta$ or from $\beta$ to $\alpha$) and in the following it will be denoted
$E_i^{e-h}$ with $i$ running from $1$ to $N+1$. Index $i$, when used to label 
a junction will refer to the one between conductors $i-1$ and $i$. 
 We will use the term {\it contact} junction for those junctions
which connect an island and an electrode, and {\it bulk} or inner
junction for those
ones in between two nanoparticles.
For the contact junctions $i=1,N+1$ $E_i^{e-h}\sim E^{isl}_c$ as $E^{source,drain}_c<<E^{isl}_c$ while for the bulk junctions
 $i=2$ to $i=N$ $E_i^{e-h}=2E_c^{isl}$.
The second
term in (\ref{change}) can be seen as the change in potential between the 
conductors involved in the tunneling. 
The potential at each site depends on the charge 
state of the array prior to the
tunneling event.
At the electrodes
$\phi_0=V_0$, $\phi_{N+1}=V_{N+1}$.
At the islands, the potential can be decomposed into two terms: 
$\phi_i^{dis}$, a random potential due to random charges in the substrate and 
$\phi_i^{ch}$ a
potential due to the charges in the islands.
$\phi_i=\phi_i^{dis}+\phi_i^{ch}$,
with
\begin{equation}
\phi_i^{ch}=\frac{Q_i}{C_i}
\end{equation} 

Analogously we can define the potential drop at each junction
\begin{equation}
\Phi_i=\phi_i-\phi_{i-1}
\end{equation}
with the corresponding disorder and charging terms $\Phi_i^{dis}$ and 
$\Phi_i^{ch}$. The potential drop at a contact junction
depends on the disorder and charge state of the nanoparticle and on 
the applied bias voltage. On the contrary the potential drop at a 
bulk junction is not affected by the bias voltage, except via a change in 
the charge state.

In the following we rewrite the potential at the electrodes as $V_0=\alpha V$ 
and $V_{N+1}=(\alpha -1)V$. The total 
potential drop through the array 
is $V_0-V_{N+1}=V$.  
In our model some measurable properties depend on the value of $\alpha$, which 
characterizes how the bias voltage is partitioned between source and drain 
chemical potential shifts. In several previous works the value of $\alpha$ was 
chosen either as $\alpha=1/2$, correspondingly to a symmetrically biased 
array, or as $\alpha=0,1$ corresponding to completely asymmetric biasing.
$\alpha=1$ has been also called the forward bias 
condition\cite{Stopa01}
Both values have been used in the literature, mostly without discussion.
In the symmetrically biased case the potential drop at both contact junctions 
is equally modified 
by the bias voltage. On the contrary for $\alpha=0 (1)$ only the drain (source)
junction is affected by the bias.   
Since no physical properties depend on the 
overall 
zero of energy, varying $\alpha$ in our model is entirely equivalent to 
rigidly shifting all impurity potentials by $- \alpha V$.
The dependence on $\alpha$ discussed below, corresponds in part to a dependence
 on 
the alignment of the equilibrium source and drain chemical potentials with 
respect to the addition and removal energies of the electron. 
For example
for a single nanoparticle on whether the chemical potential shift required to 
add or remove 
an electron is larger.  Since in our model all transport occurs by 
transfer between adjacent nanoparticles, the evolution of a nanoparticle array 
as the bias voltage is applied is sensitive to $\alpha$.  For a given 
nanoparticle 
array with a fixed set of disorder potentials, we believe that the dependence 
on 
$\alpha$ discussed below should in principle be observable.

Whenever not specified we assume that all the junction resistances $R_i$ 
are equal 
and given by $R_T$. The effect of non homogeneous resistances will be studied 
in two ways. One of the junction resistances at a given position is larger than
 the other ones (given by $R_T$) or resistances, varying in between two values 
are randomly assigned to the junctions. To mimic
that disorder in resistances originates in variations in distances between the
islands and the exponential dependence of the junction resistance on the 
distance between islands the 
junction resistance is given by  $R=R_0 exp(\gamma dist)$ with $R_0$ and 
$\gamma$ input parameters and $dist=1 + random/2$. Here $random$ is a random 
number between $0$ and $1$. In the paper, we have used $R_0=1.1825 R_T$ and 
$\gamma=1.526,1.95,2.84$. With these values the resistance changes 
respectively 
between (5-11)$R_T$, (8-21)$R_T$ and (23-83)$R_T$.

\section{THRESHOLD VOLTAGE}
In this section we analyze the dependence of the threshold voltage on the
array parameters. 
The threshold voltage is controlled by changes in energy in tunneling and
not affected by the resistance of the junctions. Thus, we do not address the
case of disorder in resistance in this section, as it does not modify the
threshold compared to the equal-resistances case.

The threshold voltage is the minimum bias voltage  at which current can flow
through the array.
A finite bias voltage can assist the entrance of charge to
the array from the leads, as it creates a potential drop at the contact 
junction, which can overcome the excitonic energy.
A finite current requires that charges are able to be transferred 
from one electrode to the other one across the entire array. If charge 
flow can occur 
between the leads, the threshold voltage is the
minimum voltage which permits the entrance of an electron or hole into 
the array. However it is possible for charge to become stacked inside 
the array due to the disorder potential configuration or due to the lack of 
potential drops across the bulk junctions. 
In this case the threshold will be controlled by the flow of charges.

Here for disordered arrays we recover previously 
predicted\cite{Middleton93}
values for the average threshold $<V_T>$ and its root mean square deviation 
$\delta V_T$. $<V_T>$ is proportional to the number of particles and 
$\delta V_T$ to $N^{1/2}$. 
On the contrary, for clean
arrays and strictly onsite interactions the threshold voltage differs from
the one expected by extrapolating to zero coupling the value obtained for 
weakly coupled nanoparticles. In particular, for finite coupling (or its zero 
coupling extrapolation) an $N$-independent threshold voltage is predicted for
large arrays, while for strictly zero coupling we obtain 
$V_T \sim 2 E_C^{isl}N$.   
We also show that its value depends  
on $\alpha$.

The threshold corresponding to the clean case is 
plotted in Fig.~1(a) for the case of symmetrically
biased arrays ($\alpha = 1/2$),  antisymmetrically biased arrays
($\alpha = 1$), and an intermediate biasing ($\alpha = 3/4$).
In the symmetrically biased case, $V_T$ shows a step-like
dependence on $N$. 
There is a clear even-odd effect. 
The even-odd effect in the threshold voltage with the number of 
particles is absent for $\alpha=1$ and $\alpha = 0$ with a threshold
voltage $V_T=E^{isl}_c(2N-1)$.  
At strictly zero temperature the tunneling rate 
$\Gamma(\Delta E)=\frac{-1}{R_T}\Delta E \Theta(-\Delta E)$ and vanishes 
when $\Delta E$ is positive or zero. 
At the inner junctions 
$\Delta E$ is independent of the bias voltage and zero or negative
if the two islands 
differ just by just a single charge. Thus the tunneling rate vanishes. 
$N-1$ junctions 
prevent the flow of charges.
A charge gradient at each bulk junction has to be created to allow
flow of charge. 
In the symmetrically biased case, increasing the potential at the electrodes 
allows positive and 
negative charges to enter from the source and the drain respectively.
These charges accumulate on the array and create potential
drops across the bulk junctions.
At voltages just below the threshold,
the accumulated charges at the first and last islands are
equal in number and opposite in sign.
Current starts to flow at 
voltages larger than $V_T=2N E^{isl}_c$ when
$N$ is odd and $V_T=2(N-1)E^{isl}_c$ when $N$ is even, 
as these values allow for the build up of  $\pm(N-1)/2$ and $\pm(N-2)/2$
charges at the first and last islands for odd and even $N$
respectively.
This corresponds to a charge gradient $dQ_{i} = Q_{i} - Q_{i-1} = -1$,
across all bulk junctions for odd $N$ 
and across all bulk junctions except one for even $N$.  
For $\alpha=0,1$, charge can only enter the
array from one lead and the energy barriers across
{\it all} $N-1$ bulk junctions must be overcome by 
accumulated charges.
When $\alpha=1/2$ the first and 
last junctions are equivalent
and charge can enter
from both leads so it is possible in some cases for the energy barrier across
one of the bulk junctions to be overcome by the potential
drop due to two injected charges of opposite sign from
the two leads, i.e. one of the junctions can be uncharged.
The absence of this possibility
is what removes the even-odd effect when 
$\alpha = 0$ and $\alpha = 1$. An intermediate situation is found for 
$\alpha=3/4$.

Complementary information can be obtained by looking at 
the threshold voltage 
as a function of $\alpha$ for different values of $N$
in Fig.~1(b).
The threshold voltage changes in a periodic 
way with $\alpha$. 
The periods
of the features in $V_T(\alpha)$ depend on the number of barriers in the
array.
Dependence of $V_T$ on $N$ and periodic features
in $V_T$ with respect to $\alpha$ reflect the number of 
charges which have to accumulate in the first and last island prior 
to current flow.  
Fig.~1(b) 
includes curves corresponding to $V_0$ and $V_{N+1}$ for specific values
of $Q_1$ and $Q_N$.
As the assymmetry of the array
increases (increasing $|\alpha - 1/2|$), the threshold voltage is alternately
determined by the cost of injecting a charge unto the array from the source
and the drain.  At all $\alpha$ except for the values that lead to the minimum
$V_T$ for a given array length, the difference in the charge occupying
the first and last islands, $Q_1 - Q_N$, equals $N-1$.  At the minimum
values of $V_T$, this charge difference equals $N-2$.

\begin{figure}
\includegraphics[width=3.5 in]{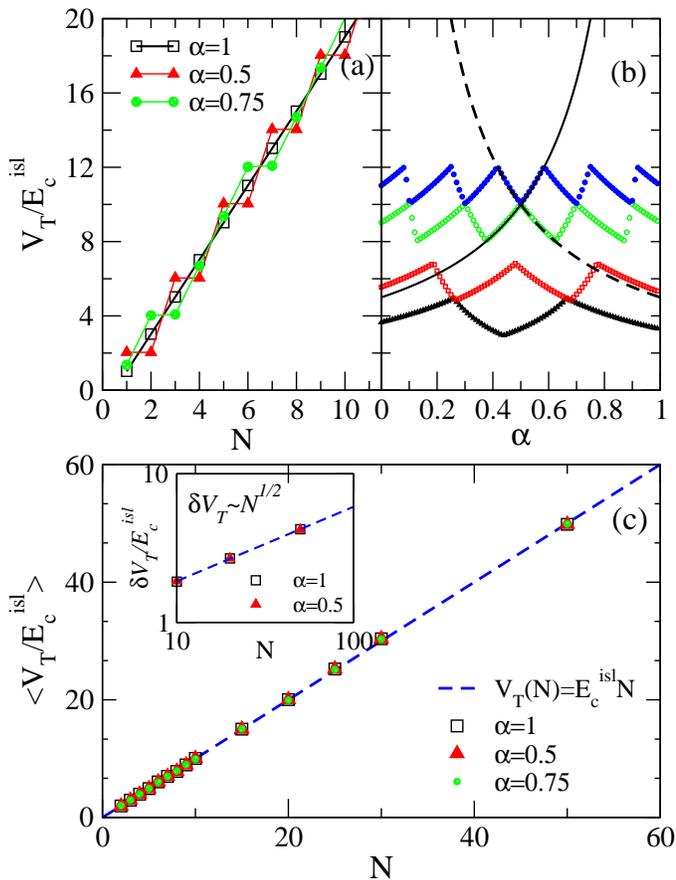}
\caption{(a) Threshold voltage of clean arrays as a function of the number of
islands $N$ for different values of the asymmetry bias parameter $\alpha$. A 
clear even-odd effect is present for symmetrically biased arrays, 
$\alpha=1/2, (V_0=V/2=-V_{N+1})$. (b)Threshold voltage as a function of 
$\alpha$.
 From top to bottom  lines with symbols correspond to clean arrays 
with 6 and 5 islands
 and to disordered
arrays with 6 and 5 islands. The peak and valley dependence reflect the change 
of contact junction which control the entrance and the number of charges which 
have to accumulate in the corresponding contact island. For comparison thin 
solid and dashed lines give, for clean arrays, the bias voltage at which 2 
and -2 charges can be 
placed at the first and last islands. (c) Main figure: average threshold 
voltage for disordered arrays as a function of the number of islands. The 
dependence of $V_T$ on $\alpha$ disappears on average and a linear dependence 
on $N$ is 
recovered. Fluctuations in threshold voltage follow $\delta V_T \sim N^{1/2}$ 
as predicted\cite{Middleton93} }
\end{figure}

The clean case of a related system was studied by Hu and O'Connell\cite{Hu94}. 
They 
analyzed a one-dimensional array of N gated junctions with equal junction 
capacitances 
$C_J$ and equal gate capacitances $C_g$.
Due to the finite value of $C_J$ charges in a given island interact with 
charges in other islands and with charges in the electrodes. With an applied 
bias voltage the interaction between charges in the electrodes and in the 
islands results in a bias induced potential drop at the bulk junctions.
Once a charge is injected unto the array, it will 
have no difficulty in traveling through it, and the threshold voltage equals
the voltage required for injection of a charge from the electrodes.  
As the ratio $C_g/C_J$
increases the threshold voltage of a long array tends to an N-independent 
value of the order of the charging energy.
The onsite case discussed here corresponds to $C_J=0$.  
If one extrapolates the case discussed by Hu and O'Connell\cite{Hu94} to
$C_g/C_J\rightarrow 0$ an N-independent threshold voltage would be expected
for onsite interactions. 
As shown above, the threshold voltage of clean arrays does not satifies this
prediction,
as at zero temperature the charges cannot
travel freely through the array, and the threshold voltage increases with
the number of islands.

In the case of disordered arrays,
each array has a threshold voltage $V_T$ dependent on 
the given configuration of disorder $\{\phi^{dis}\}$.
The threshold voltage depends on $\alpha$, in a way similar
to the clean case. see Fig. 1(b). However, as shown in Fig. 1(c), this
dependence on $\alpha$ disappears in the average value and we recover
Middleton and Wingreen prediction\cite{Middleton93}.  
For the disordered case
Middleton and Wingreen\cite{Middleton93} predicted a 
linear dependence of  the threshold 
voltage on the array length. Upward steps in the
disorder potential $\Phi_i^{dis} > 0$ prevent the flow of charge.
The downward steps $\Phi_i^{dis}<0$ facilitate it.  
In average there are $N/2$ upward steps. To overcome such steps
a charge gradient has to be created in those junctions. For onsite
interactions this results in\cite{numeron} 
$< V_T>=E_c^{isl} N$. 
In this limit, they also argued that fluctuations in the
values of $V_T$ over many random configurations of disorder
 are analogous
to the fluctuations in the net distance traveled by 
a 1-D random walk with $N$ steps, resulting in $\delta V_T \sim N^{1/2}$.

Compared to arrays with no disorder, the threshold of
disordered arrays in the onsite limit tends to be smaller because in 
the clean case, all the bulk junctions inhibit
charge flow whereas in the disordered case, only
the bulk junctions with positive $\Phi_i^{dis}$
block charge flow.  
As shown in the inset of  Fig. 1(c),
we also recover the relationship for
the fluctuations in the threshold predicted by
Middleton and Wingreen\cite{Middleton93}, $\delta V_T \propto N^{1/2}$.

\section{FLOW OF CURRENT}
For bias voltages larger than threshold
the current $I$ can flow, but it is a strongly non-linear function of voltage. 
The 
current depends on  the charging energy and number of islands, 
the presence or not of charge 
disorder in the array, the resistances of the junctions and on the asymmetry 
of the applied bias voltage.   
Linearity and independence on $\alpha$ is recovered at large voltages. As 
discussed by Middleton and Wingreeen\cite{Middleton93} a power-law dependence 
of the 
current on voltage is found close to threshold. In this section we discuss
the different regimes which can be differentiated in an I-V characteristics and
its dependence on the input parameters.  At very low voltages by comparing 
analytical and numerical results we resolve the
controversy on the exponent of the power-law and show it to be linear with
a slope which depends on $\alpha$ and the resistance of the contact junctions, 
but that is independent on the array length (except in a particular case in 
which an even-odd effect is found). The linearity is however restricted to
very small values of $V-V_T$.
We also clarify the dependence of the
Coulomb staircase profile on the bias parameter $\alpha$ and estimate the
asymptotic current at high voltages, and the bias voltage at which this
high voltage regime is found. Very large values of the bias voltage have to 
be applied to reach this linear dependence. 

\subsection{Linear dependence close to threshold}

There has been some controversy regarding the power-law of the current 
with $(V-V_T)$ through one dimensional disordered arrays 
for voltages close to $V_T$, $I \sim (V-V_T)^\nu$. 
Middleton and Wingreen\cite{Middleton93} 
predicted linear behavior for both the long and short range interaction. 
Reichardt and Reichardt\cite{Reichhardt03} found a square root 
behavior using a model with a $1/r$ interaction between the charges in 
the islands.  They argued that $\nu=1/2$ is the 
exponent 
corresponding to an sliding 
charge-density wave. They pointed out that the larger values of the exponent 
obtained by 
Middleton and Wingreen\cite{Middleton93} are a consequence of using voltages 
which 
are not small enough. 
Kaplan et al.\cite{Kaplan03} found $\nu=1$ in the long-range limit of an 
array of 
dots capacitively coupled to their nearest neighbors. 
Finally Jha\cite{Jha05} and Middleton argued that the dependence of the 
current 
of disordered arrays in the onsite limit on $(V-V_T)$ for voltages 
marginally greater than $V_T$ is linear with an slope inversely proportional
to the length of the array. Numerically they found an approximate linear
behavior only in the 
case of very long arrays but not for the smallest voltages analyzed (where 
they found a sublinear dependence) but in an intermediate voltage regime.  
So far, there are no experiments available in  completely 
one-dimensional arrays, but there are a few in quasi-one dimensional systems.
The approximate power-law measured\cite{Tinkham99} at voltages 
$(V-V_T)\sim 0.01 V_T$ is larger than 
unity which has been attributed to the fact 
that the system is not strictly one-dimensional.

In this subsection, we 
show that the current varies linearly with respect to $V-V_T$ for very small 
$V-V_T$ 
and that the slope is not inversely proportional to $N$.  The slope is 
sensitive to the resistance of the contact junctions and to the degree of
symmetry between the applied bias voltages on the source and drain
leads.
While this linear dependence is found analytically and in the numerical
calculations  
the value of $(V-V_T)$ at which it disappears is  
very small. Because of the small value of the current at such voltages, the 
linear dependence most 
probably cannot be seen experimentally. 

\begin{figure}
\includegraphics[width=3.3 in]{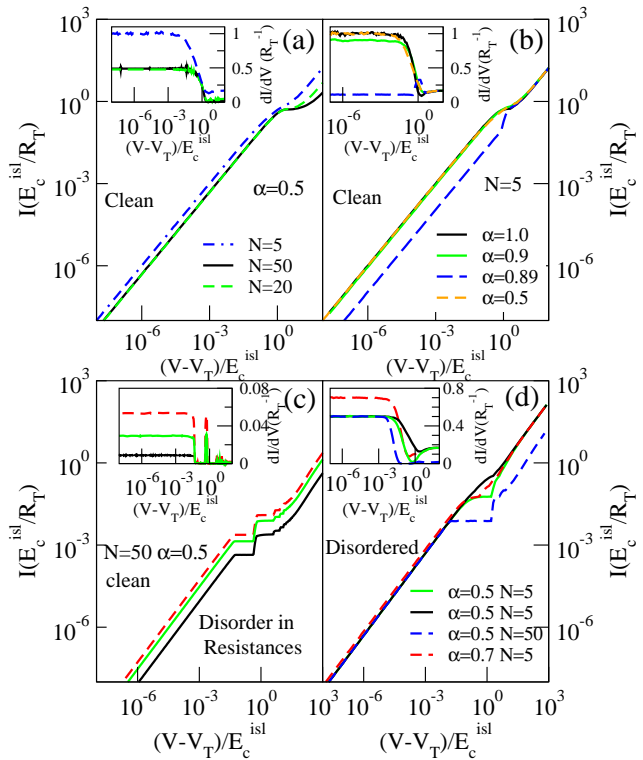}
\caption{Main figures:
I-V curves in logarithmic scales for different array parameters. Insets show the derivatives 
(in units of 
$1/R_T$) of the curves plotted in the main figures.The linear dependence of 
current on 
voltage (constancy of the derivative) is clearly seen in all the plots but it disappears for 
$V-V_T \sim 10^{-2} E_c^{isl}$
which for the cases shown corresponds to $(V-V_T)/V_T \sim 10^{-4}$. 
(a) to (c) show I-V curves corresponding to 
arrays without charge disorder. All junction resistances are equal in (a) and 
(b). As shown in (a) for clean arrays
the slope of the linear 
dependence does not depend on the number of islands except for $\alpha=0.5$ 
which shows an even-odd effect with a slope equal to unity and $0.5$ for odd 
and even number of particles, respectively. (b) The dependence of slope on 
$\alpha$
 can be non-monotonous if there is a change in the contact junction which acts 
as a bottle-neck. (c) I-V curves for arrays with junction resistances randomly assigned varying 
between $(5-11) R_T$ (upper curve), $(8-21) R_T$ (middle curve) and 
$(23-83)R_T$ (bottom
 curve). The 
different slopes are due to different resistances  at the
bottle-neck junction. (d) I-V curves of disordered arrays with homogeneous 
contact resistance. The $\alpha=0.5,N=5$ curves correspond to different 
realizations of disorder. The even-odd effect present for $\alpha=0.5$ 
in the clean case has 
disappeared as just one contact junction acts as bottle-neck.}
\end{figure}

The linearity above but very close to threshold can be easily understood.
Charges can enter only through the contact junctions from the leads. 
The current through the array is equal to the average charge transferred per 
unit time. The average time necessary to transfer a charge through the array 
is the sum of the time involved in all the processes in the sequence of 
tunneling events from the 
moment in which charge enters the array from one electrode until it leaves the 
array to the other one. 
If the time 
associated to tunneling at a given junction is much larger than the 
time involved in the rest of processes, this junction  acts as bottleneck 
and the time
necessary to transverse the array is approximately 
equal to the inverse of the
scattering rate through this junction. 
Thus, the current can be approximated by the tunneling rate 
across the bottle-neck 
junction. 
Below 
threshold, but close to it, the tunneling at the contact junctions costs 
finite 
energy and transport is suppressed. This cost in energy is reduced by the
applied bias voltage 
and at threshold is zero at the entrance junction. When $V >V_T$ but very 
close to 
it, 
the junction from which the charge enters the
array acts as bottle-neck for transport, once it is allowed. 
At zero temperature, the tunneling rate is $-\Delta E/R$.
The dependence on bias voltage of the energy for tunneling compared to 
the one at threshold,
at the source and drain junction is, 
respectively $\delta E_{S}=-\alpha (V-V_T)$ and 
$\delta E_D=-(1-\alpha)(V-V_T)$. 
If charge enters
through a single junction, the current approximately equals 
\begin{equation}
I=\frac{1}{R_1}\alpha(V-V_T)
\label{primeq}
\end {equation}
if charge enters from the source, and
\begin{equation}
I=\frac{1}{R_{N+1}}(1-\alpha)(V-V_T)
\label{segundaeq}
\end{equation}
if holes enter unto the array from the drain.
Both source and drain junctions have to be taken into account in 
the clean symmetrically biased case when $N$ is odd.
\begin{equation}
I=\left (\frac{1}{R_1}+\frac{1}{R_{N+1}}\right )\frac{1}{2}(V-V_T)
\label{ultimaeq}
\end{equation}
When $N$ is even and $\alpha=0.5$ it is necessary that charge enters through 
both junctions for 
current to flow and current is approximately equal to
\begin{equation}
I=\frac{2}{R_1+R_{N+1}}\frac{1}{2}(V-V_T)
\label{nuevaeq}
\end{equation}
This behavior is observed in Fig.~2. The linear behavior is clearly appreciated
in both the log-log scale in which the main figures are plotted as well as
in the constancy of the derivatives in the insets. The dependence of the
slope of the I-V curves is better seen in the conductance plotted in the
insets. As seen in (a) the $dI/dV$ is equal for the $\alpha=0.5$, $N=50$ and 
$N=20$ curves, 
i.e. independent on the array length. On the contrary, it is double for $N=5$.
This behavior originates in the even-odd alternancy predicted by 
Eqs.(\ref{ultimaeq}) and (\ref{nuevaeq}). The dependence on $\alpha$ is studied
in Fig.2(b). From top to bottom, the change of slope with $\alpha$ is smooth 
and given by $\alpha/R_1$ 
as predicted by (\ref{primeq}), but  turns non monotonously if currents 
starts being controlled by
(\ref{segundaeq}) with slope $(1-\alpha)/R_{N+1}$. As also expected from the
discussion above the slope is affected by disorder in resistances in Fig.~2(c),
 but not by charge disorder in Fig.~2(d), except in the lose of the even-odd
effect present in the clean case for $\alpha=0.5$.

 
\subsection{Loss of linear behavior and intermediate regime}

The linear behavior in Fig. 2 appears for several orders in magnitude. In
spite of this, it disappears for $V-V_T\approx 10^{-2}$ or $(V-V_T)/V_T \sim 10^{-4}$.
The magnitude of the 
current is probably too small for this linearity to be detected experimentally.
We have found that linearity gives rise to sublinear behavior when 
the time of the other tunneling processes  become relevant 
compared to the time spent at the bottle-neck.
Upon increasing $V$ the tunneling rates of the different
processes involved in the transport become more homogeneous. 
To obtain sublinear behavior, 
it is just necessary than the two slowest processes in a sequence have 
comparable rates.   
The lost of linearity can depend on the resistance of the junctions when a 
non bottle-neck junction has a resistance much larger than the bottle-neck one.
The bottle-neck character of a process at a junction at the leads can
disappear faster for longer arrays as there are more tunneling processes which 
will contribute to the total time. In the disordered case, the energy gain 
of some of the tunneling processes is smaller than in the clean case 
and the contact junctions can stop being the bottle-neck earlier, i.e. for 
smaller $V-V_T$. But, in general we have not found very significative 
differences for different array parameters in the value of the bias voltage 
at which the linearity disappears. 

As the energy of a tunneling process through an inner 
junction does not depend on the applied voltage (except via the charge 
accumulated on it), when a bulk junction controls the transport, the current 
is independent on voltage showing a characteristic staircase profile.
The existence of the Coulomb staircase has been known for a long 
time\cite{singlecharge,averinlikharev,amman91,hanna91}. Early 
claims 
reported a Coulomb staircase only in the asymmetrically biased 
case\cite{singlecharge}. 
More recent results in clean capacitively coupled nanoparticle arrays, show 
that a 
staircase also emerges in a symmetric array under symmetric 
bias\cite{Stopa01}, 
but claim that the I-V characteristic for an N-dot array under forward bias is 
identical to that for a 2N-dot one under symmetric bias. We show here that 
while the appearance of the staircase is generic, the last statement is not 
correct.

The current has kinks at those voltage values which change the maximum 
number of
charges which can be accumulated at the first or last island, allowing new 
transport processes. These voltage values
depend on  the asymmetry $\alpha$ and the existence of charge disorder in the 
array  but not on resistance disorder. To allow the addition of an extra
charge in the first or last nanoparticle requires an increase in bias voltage
in the adjacent electrode of approximately $2E_c^{isl}$. In the clean case, 
when $\alpha=1,0$ only one electrode changes its potential and the width of the
steps in bias voltage is $2E_c^{isl}$. On the other hand, when $\alpha=1/2$ 
the 
change
in potential of a given electrode is just the half of the bias voltage and 
steps appear in intervals of $4E_c^{isl}$. 

With charge disorder and $\alpha=0,1$ the position of the kinks slightly
depends on voltage, but the width of the voltage intervals between the kinks
does not change, as new charges are added 
through a single junction. If $\alpha=1/2$, charges enter from both contact 
junctions but the corresponding kinks in the current do not appear at the same
position. While the width of a kink corresponding to a given junction remains 
$4E_c^{isl}$ in a general case in the I-V characteristic 
there will be two kinks in each $4E_c^{isl}$ interval in 
bias voltage due to the alternative position of the kinks of both contact 
junctions. Except in very special cases the separation of kinks does not do 
equal 
$2E_c^{isl}$. 

The position of the kinks in the clean and disorder cases can be 
observed in Fig.~3. Two main features can be observed. For clean arrays 
(insets in (a) and (b)) at the
onset the current shows a big jump. Once the charge gradient is created and 
charge can transverse the array, it flows easily. The steps at higher voltages 
are
much smaller but have the width predicted above. In the main figures 
corresponding to disordered arrays, the large big jump has disappeared and  
steps are more clearly observed. Two stepwidths, which add $4E_C^{isl}$ are 
seen for $\alpha=0.5$ and just steps with width $2E_C^{isl}$ appear for 
$\alpha=1$.

\begin{figure}
\includegraphics[width=3.3 in]{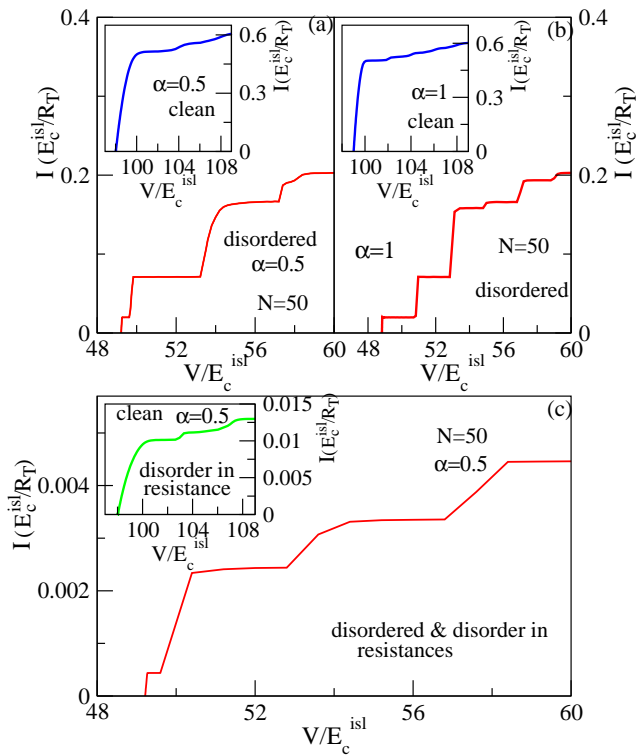}
\caption{I-V curves for N=50 and different array parameters  at intermediate
 bias voltages showing the Coulomb staircase. Insets in (a) and (b) 
correspond 
to an array without charge or resistance disorder and respectively
$\alpha=0.5$ and 
$\alpha=1$. The current shows a big jump at threshold and very weak small 
steps at 
higher voltages.The width of the steps is $4E_c^{isl}$ in (a) and $2E_c^{isl}$ 
in (b). The staircase structure is completely washed out with increasing bias 
voltage. 
Main figures in (a) and (b) show I-V curves for arrays with charge disorder 
but homogeneous resistances. $\alpha=0.5$ in (a) and $\alpha=1$ in (b). The 
first step height is reduced and the 
staircase structure is more pronounced than in the clean case because tunneling
 processes in the bulk  can
have small energy gain and become the bottle-neck more easily. The width of the
steps in (b) remains equal to $2E_c^{isl}$ as in the clean case. In (a) there
are two different step lengths which alternate and sum to $4E_c$, 
corresponding to
the entrance of charge through both electrodes.
The I-V curves in (c) correspond to arrays with resistance disorder and 
$\alpha=0.5$ without charge disorder in the inset, and with charge disorder in 
the main figure. Disorder in resistance in general favors  that bulk junctions 
act as 
bottle-neck for the current, but for $\alpha=0.5$ it affects differently to 
the steps associated to any of the contact junctions.}
\end{figure}

Previously, in the analysis of the threshold voltage, we saw that while
the value of $\alpha$ has an effect the value for the onset of current, 
it does not 
seem possible to determine $\alpha$ from a measurement of $V_T$ in the, 
experimentally relevant case, of a disordered array. However if the onsite 
interaction case discussed here can be experimentally reproduced, 
the width of the steps in 
the I-V curve can differentiate $\alpha$ values.  
Disorder in the resistances does not modify the voltages at which kinks in the
current appear but it does affect the staircase profile.
The staircase
profile is modified in Fig.~2(c), compared to (a) due to the disorder in 
resistances.
A very large 
resistance in a bulk junction can sharpen the steps, as it creates a 
bottleneck 
for the current at a junction with an associated energy for tunneling which 
does not directly depend on bias voltage, but the opposite behavior can also
take place if the large resistance if found at any of the contact junctions. 
Note that the particular way in which the I-V curve is affected by
disorder in resistance depends on the particular resistance and 
charge-disorder distributions.

\subsection{Linear Regime at High-Voltages} 

At larger voltages linear dependence is recovered. The asymptotic linear 
I-V  does not
extrapolate to zero current at zero voltage, but it cuts the zero current axis
at a finite offset voltage. 
This high-voltage linear regime can be understood 
analytically.
We obtain that the slope of the I-V curve depends only on the sum of 
the junction resistances in series and the offset voltage is given by the sum 
of the excitonic energies of all the junctions. Contrary to what was found at 
low voltages, close to threshold, the offset voltage for purely onsite 
interactions recovers the 
zero inter-island capacitive coupling value\cite{Schon,Bakhalov91} calculated 
starting from a finite value of the inter-island capacitance.
To assist comparison
with experiments we compute the voltage at which this linear regime is
reached. 

\begin{figure}
\includegraphics[width=3.3 in]{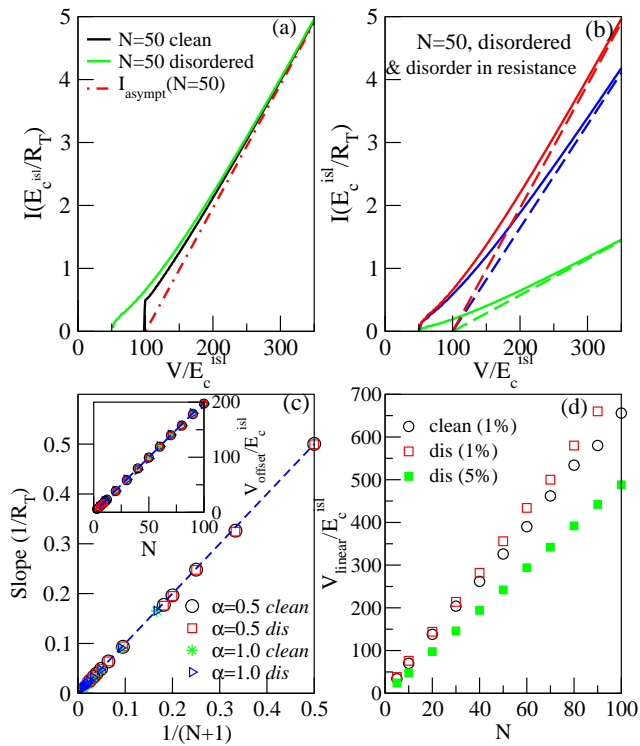}
\caption{
(a) I-V curves calculated up to high-voltages for $N=50, \alpha=0.5$ for clean
and charge-disordered arrays. Both curves approach the same asymptotic curve at
high voltages, even if the threshold voltage is quite different. Theoretical 
prediction is included for comparison.(b) Computed I-V curves for 
charge-disordered arrays with different junction resistances (solid lines) 
with their theoretical asymptotic predictions (dashed lines). From top to 
bottom a curve corresponding to an array with all-equal junction resistances, 
an array with randomly assigned resistances and an 
array with all-equal random resistances, except the first one which
is ten times $R_T$. The slope differs but all the curves have equal offset 
voltage. 
(c) Slope (main figure) and offset voltage (inset) which give better fitting 
to the numerically computed current at high-voltages as a function of the 
number of islands in the array, for several array parameters, all of them
with homogeneous junction resistances. For comparison the 
theoretical prediction (dashed-line) is included. The offset voltage is 
proportional to the number of islands and given by $\sum_{i=1}^{N+1}E_i^{e-h}$,
 while the slope goes like the inverse of the sum of the junction resistances 
added in 
series. It is inversely proportional to the number of junctions, when 
all resistances are equal. Independence on the value of $\alpha$ and the 
presence or absence of charge disorder is observed.
(d) Voltage at which the high-voltage asymptotic behavior is reached, 
estimated as the value at which $I-I_{asympt}/I$ is smaller than a given value,
 1\% and 5\% in the figures.It is slightly larger for disordered arrays, 
increases linearly with the number of islands and 
it is approximately three and 2.5 times the offset voltage. In large arrays a 
very large bias voltage can be required to reach this linear dependence.}
\end{figure}

At very high 
voltages, the charge gradient ensures that all the tunneling processes 
to the right decrease the
energy. The corresponding tunneling rates are 
$\Gamma_i=R^{-1}_i(\Phi_i -E^{e-h}_i)$ and the total tunneling rate 
$\Gamma^{tot}$ for no resistance disorder is
$\sum_{i=1}^{N+1}\Gamma_i$. Having in mind that $\sum_{i=1}^{N+1}\Phi_i=V$, 
$\Gamma^{tot}=R^{-1}_i\left ( V-\sum_{i=1}^{N+1} E^{e-h}_i\right )$. This rate 
is independent on the selected tunneling process.
To transfer a charge from the source to the drain requires in average $(N+1)$ 
tunneling events. The average
current is thus
\begin{equation}
I_{asympt}\sim \frac{1}{(N+1)R_T}\left ( V - \sum_{i=1}^{N+1}E^{e-h}_i \right )
\end{equation}
This prediction is compared in Fig.~4 (a) and (c) with numerical results. 
The slope of the current does not depend on  
$\alpha$ or the existence of charge disorder, but only on the number of 
junctions 
$(N+1)$. The slope is the same as obtained adding all the resistances in 
series. Asymptotically, at low voltage, this curve
cuts the $I=0$ axis at the offset voltage 
$V_{offset}= \sum_{i=1}^{N+1}E^{e-h}_i$. This value is, 
in general, different 
to the threshold voltage and independent of
the resistance of the junction and the asymmetry of the bias voltage. 

Previous derivation is valid even if there is inhomogeneity in the value of 
island capacitances but 
relies on the homogeneity of the junction resistances through the 
array. If this is not
the case the total tunneling rate of each step in a sequence is 
$\Gamma^{tot}=\sum_{i=1}^{N+1}R_i^{-1}(\Phi_i-E_i^{e-h})$.
One could argue that on average, the charge gradient would be such that it 
ensures a uniform tunneling rate
$\Gamma^{uni}$ through  all the junctions. 
The potential drop which gives such 
a tunneling rate is 
$\Phi_i=R_i\Gamma^{uni} +E^{e-h}_i$ and 
$\Gamma^{uni}=\left( V-\sum_{i=1}^{N+1}E^{e-h}_i\right )/R_{sum}$ with 
$R_{sum}=\sum_{i=1}^{N+1}R_i$. 
There are $(N+1)$ possible tunneling events at each 
step 
in a sequence and $(N+1)$ steps, thus both $(N+1)$ factors cancel our. 
The resulting average current is
\begin{equation}
I_{asympt} \sim \frac{1}{R_{sum}}\left (V-\sum_{i=1}^{N+1}E^{e-h}_i\right )
\label{currenthigh}
\end{equation}
As in the uniform resistance case, the slope in the current corresponds 
to the addition in 
series of all the resistances. The predicted
asymptotic high-voltage behavior is observed in Fig. 4 
for arrays with different
parameters.
 Note that as longer is the array 
larger voltages have to be applied to reach this voltage. The voltage 
$V_{linear}$ at which
the linear behavior is reached is estimated in Fig.~4(d). It is approximately 
three times the offset voltage and slightly larger in 
the presence of charge 
disorder. In long arrays $V_{linear}$ can become very large and the linear 
behavior will not be easily reached experimentally. 

\section{Potential drop through the array}

The potential through the array can nowadays be measured\cite{Sample02}, but 
to our knowledge
it has not been studied theoretically. 
In conventional ohmic systems  with a linear current-voltage relation $V=IR$ 
the 
potential drops 
homogeneously through the array if the resistivity of the system is 
homogeneous. When the proportionality constant between voltage and current 
is given by the sum of the resistances in series, but these resistances are not
all equal, the voltage drop at each point is proportional to the local 
resistance. 
The nanoparticle array I-V characteristics are highly non-linear
and in general it is not obvious how the potential drops through it. 
At the islands the potential is the sum of the disorder and charge terms, while
at the electrodes the potential  is controlled by the applied bias. 
As seen 
above, there are  two linear regimes which can be identified. 
At high voltages the differential conductance equals the inverse of the 
sum of the resistances in series. 
Naively, a potential drop proportional to the resistance at each junction
could be expected at these voltages, but we will see that this is not
exactly the case.
In the
low-voltage regime the slope of the current does not correspond to the 
addition of the resistances in series, but it is proportional to a single one 
(to the sum of two of them in the symmetric case) and it is not clear that 
the  
potential drop should be proportional
to the junction resistance. 
In this section we study the
potential drop through the array at low, intermediate and high voltages 
and show that in none of these regimes the potential drop at a given junction
is strictly proportional to its resistance. In the linear regimes we show
that deviations from this proportionality are related to the offset voltages 
and filling of the array.
In the intermediate regime the voltage drop oscillates with position and 
reflects correlations between the charges.  

\subsection{High-Voltage Regime}
We start with the high-voltage linear regime, as it is the easiest to 
understand.
Fig.~5 shows the average potential drop in a clean  
and a disordered array for 
a given 
bias voltage in the high-voltage linear regime. All the junction resistances 
are 
equal in the top figures. The average voltage drop is equal in both the clean 
and disordered 
case,
and at first sight it seems linear. A linear potential drop through the array
implies a homogeneous average junction potential drop$\bar\Phi_i$.
However, at the contact junctions $\bar \Phi_i$ is approximately $E_c^{isl}$ 
times smaller than at the bulk junctions.
The voltage drop at each junction is not equal to the current divided by the
junction resistance either, as could be naively expected. The reasons for these
deviations can be found in the offset voltage in the I-V curve 
and in Eq. (\ref{change}) which gives the change in energy for tunneling.

At high voltages the I-V curve is linear but the total voltage drop through 
the array does not equal $R_{sum}I$. As, seen in equation (\ref{currenthigh}), 
there is an 
offset voltage $V_{offset}=\sum_{i=1}^{N+1} E_i^{e-h}$. This offset voltage 
reflects the excitonic energy cost for tunneling. The excitonic energy is not
equal at each junction. It is $2E_c^{isl}$ at the bulk junctions and 
approximately 
$E_c^{isl}$ at the contact ones. Only the extra potential drop 
$\bar\Phi_i-E^{e-h}_i $ at each junction 
gives a finite contribution to current through it. On average  
\begin{equation}
I=\frac{1}{R_{sum}}\left (\bar \Phi_i -E^{e-h}_i \right )
\end{equation}
From current conservation at high voltages the average potential drop through 
the  array
\begin{equation}
\bar \Phi_i=E^{e-h}_i +\frac{R_i}{R_{sum}} \left (V-V_{offset}\right )   
\label{potdrophigh}
\end{equation}
It is not affected by the presence of charge disorder in the array (but it 
would change if capacitances are not homogeneous, via $E^{e-h}_i$). As 
observed in Fig.~5, Eq (\ref{potdrophigh}) gives a good estimate of the 
potential drop. 
The validity of Eq(\ref{potdrophigh}) is better seen  
when $ \bar \Phi_i -E^{e-h}$ is plotted. It
is proportional to the resistance of the junction and equal in every junction 
if all resistances are the same. This statement is valid independently on the
position of the resistance, as shown in the bottom figure of Fig.~5 and its 
inset and on the asymmetry
$\alpha$ of the applied voltage (not shown).
The dependence of $\bar \Phi_i$ on the junction resistance is easily 
understood.
The tunneling 
probability through a junction depends on its resistance. It is inversely
proportional to it.
When the resistance is very large, the charge 
has a lesser tendency to jump from an island to its neighbor and it will spend 
more time in the island producing a dependence of
the time-averaged potential drop on the junction resistance distribution.
 
\begin{figure}
\includegraphics[width=3.3 in]{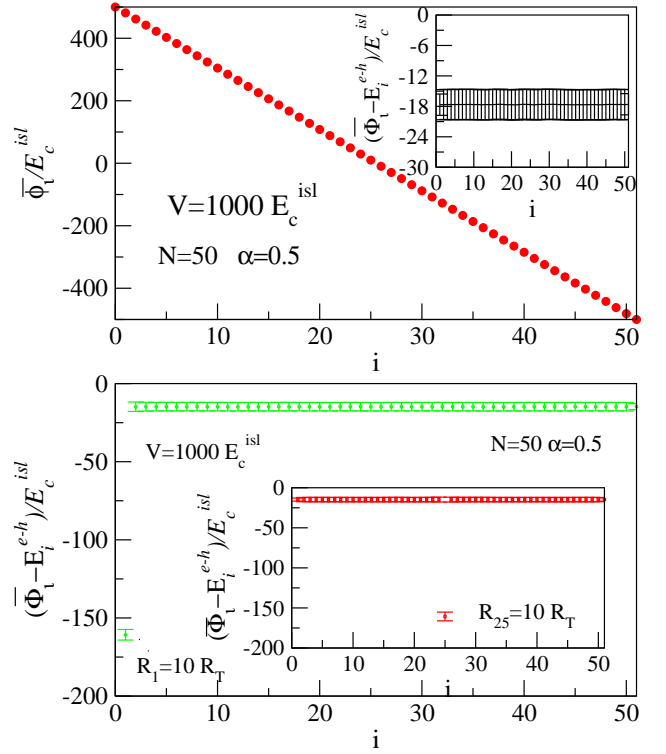}
\caption{Top: mainfigure shows the average potential at the islands 
$\bar \phi_i$ as a 
function of 
position for a 
disordered array with $N=50$, $\alpha=0.5$ and all junction resistances equal. 
The average potential drop through the array is close to be linear. As shown 
in the inset the average potential drop at the junctions $\bar \Phi_i$ is 
homogeneous only once the
excitonic energy is substracted. The value subtracted is smaller at the contact
junctions where $\bar \Phi$ is smaller. Error bars give an estimation of 
the fluctuations of the potential drop. 
Bottom: main figure (inset) show the average potential drop, with error bars 
giving its root mean square, at the junctions with 
the excitonic energy 
substracted corresponding to a clean array with the first (middle) junction 
ten times larger 
than the rest. $\bar \Phi_i-E_i^{e-h}$ is proportional to the junction 
resistance. Note that this proportionality holds only once the excitonic energy
is substracted. 
}
\end{figure}

As seen above, in this high-voltage linear regime
the current can be obtained from the average tunneling rate and correspondingly
from the average potential drop. Deviations of the average value 
$\delta \Phi_i$, i.e. the root mean square (r.m.s.), are shown in the 
inset in Fig.5, in the form of error bars. They are 
slightly smaller at the contact 
junctions as the potential at the electrodes is restored to its nominal value 
via a battery prior to any tunneling event and larger at junctions with
a larger resistance. Fluctuations in the local voltage drop $\Phi_i$ increase
with applied bias voltage as the number of possible charge states and the width
of the distribution of hopping energies do. Fluctuations are larger at those
junctions with a larger resistance, but $\delta \Phi_i/\bar \Phi_i$ is 
smaller.   

\subsection{Low-voltage regime}

Close to threshold the current depends linearly on $V-V_T$. Here we show that
in this linear regime the average potential drop mainly reflects the 
charge state of the array at threshold. 
This charge state depends on the asymmetry of the 
voltage drop $\alpha$ and disorder and in a symmetrically biased clean array on
the even or odd number of islands. For $\alpha=1$, charges enter from the 
source and  $(N-1)$ bulk junctions prevent charge motion. If an electron 
reaches the
last nanoparticle, it can freely jump onto the drain at zero potential. There 
is 
no charge gradient at the drain junction. Consequently, the potential drop at 
this junction vanishes at threshold. On the contrary, at the $(N-1)$ bulk 
junctions there is a charge gradient equal to unity, with the corresponding
potential drop $2E_c^{isl}$. To allow current $V_0-\phi_1$ equals the 
excitonic 
energy of the first junction, approximately equal to $E_c^{isl}$. Close to 
threshold, as the bottle-neck for the current is the entrance of electrons 
from the source
the charge state of the array is most of the time equal to the one at 
threshold, and only perturbed by the fast passage of charges. The average 
potential drop, plotted in Fig.~6(a) for a clean array with  
all junction resistances equal is  almost the same as the static potential
drop at threshold. 

\begin{figure}
\includegraphics[width=3.2 in]{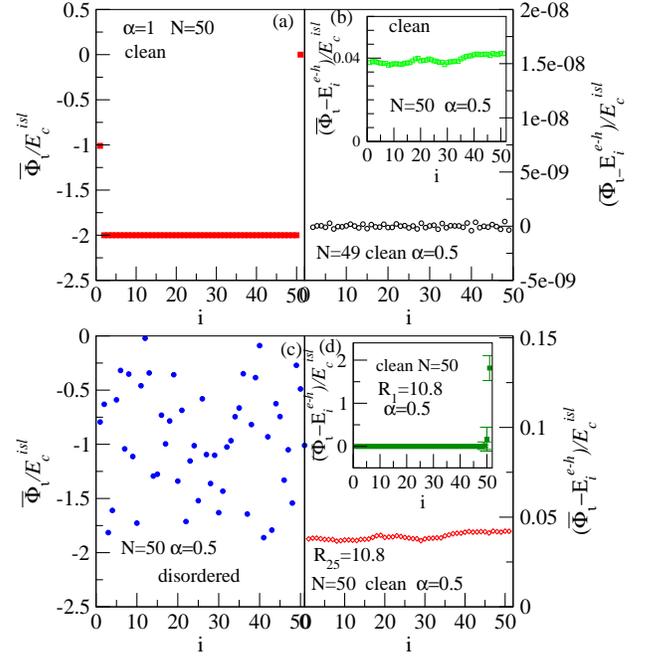}
\caption{Average potential drop through the array at bias voltages very close
to threshold. Fluctuations are smaller than the symbols. 
(a) Clean N=50 array with $\alpha=1$. 
Notice that the potential
drop vanishes at the last junction, as the source is at zero potential and
there is no charge gradient at this junction. At the first junction is
equal to the excitonic energy, which at a contact junction equals 
$E_c^{isl}$. (b) 
Average potential drop for a symmetrically biased N=49 array in main figure 
(N=50 in the inset) with the excitonic energy substracted. Once the excitonic 
energy, associated to the charge 
gradient at threshold has been substracted the average potential drop almost
vanishes for $N=49$. The homogeneous and positive value for $N=50$ reflects 
that
every junction is uncharged with equal probability.    
(c)Average potential drop at the junctions $\bar \Phi_i$ 
corresponding to a disordered array with $N=50$ islands and $\alpha=1/2$. (d) 
Main figure (inset) $\bar \Phi-E^{e-h}$ corresponding to a $N=50$ clean array 
with 
the middle (first) junction resistance 10.8 larger than the other ones. When 
the
larger resistance is in the middle its effect on the average potential drop at
these small voltages is barely visible and the potential drop almost equals 
the one found in the inset in (b). This is the generic behavior found
with resistance disorder very close to threshold. A special case is a 
N-even clean $\alpha=0.5$ array with the
first resistance larger than the other ones, shown in the inset. The average 
potential drop differs considerably with respect to the one found in the inset 
in (b). The presence of the larger resistance at a contact junction modifies 
the average charging of the array and selects the opposite contact junction 
as the one which lacks charge gradient.  
}
\end{figure}

For $\alpha=1/2$ and odd number of particles, at threshold the charge gradient
and charge potential drop in a clean array are respectively one and 
$2E_c^{isl}$ at the bulk 
junctions. At the contact junctions the potential drop is $E_c^{isl}$. As in 
the
$\alpha=1$ case discussed above, the average 
potential drop in the linear regime close to threshold,  is 
very close to the one found at threshold, which equals the excitonic energy at
each junction, $\bar \Phi_i-E^{e-h}_i \sim 0$, what can be seen in the
main figure in Fig.~6(b). When the number of particles 
is even, the charge gradient at one of the junctions vanishes. As shown in 
the inset of Fig.~6(b),
$\bar \Phi_i-E^{e-h}_i$ in the bottleneck regime is positive and equal for 
all the junctions. This reflects that every junction is uncharged with equal 
probability.

In the disordered case, only those junctions with upward steps in the disorder
potential are charged, and this is reflected in the average potential drop, 
in Fig.6 which adds disorder, charge and bias potential. 

The threshold voltage does not depend on the resistance of the junctions, but
the flow of charge does. This  is reflected on the average potential drop
on such very small scale that even if the threshold voltage potential drop is
substracted at each junction and for reasonably large changes in resistance it 
is not visible (see main figure in Fig. 6(d). This  is different  to the 
dependence 
observed in the 
high-voltage regime. For extremely large values of the
 resistance disorder a weak effect on the average voltage at bias close to 
threshold can be seen (not shown).  In this case,  the potential drop at a 
junction with a larger resistance is slighltly larger than at the rest. At the 
adjacent junctions it is slightly smaller, what reflects the average charge 
state of the nanoparticles joined by the large resistance. 

The r.m.s of the junction potential drop at low bias voltages are very small in
some of the cases analyzed, of order $10^{-3}E_c^{isl}$ in main figures in 
Fig. 6(a) and (b). 
Error bars are small 
because most of the time the array is at the threshold charge state.
This is not the case in the inset in 
6(b) where fluctuations are or the order of $0.3 E_c^{isl}$, or in the inset 
in 
6(d) where they are small everywhere except at the last two junctions where 
they are of order $E_c^{isl}$. Disorder also increases the fluctuations in the
average voltages to values of the order of the charging energy. 

\subsection{Intermediate Voltage Regime}

  The most interesting regime to analyze the voltage drop is at intermediate
voltages where the I-V curve show the Coulomb staircase. For the case of a 
clean array with capacitively coupled nanoparticles Stopa\cite {Stopa01} showed
that the steps in the I-V characteristic correspond to alternation of the 
charge density between distinct Wigner crystalline phases. In our case, the
interaction is short-range, but the possibility of such a Wigner state with 
charges
periodically ordered to minimize their repulsion, if present, 
could be observed via the potential drop. Periodic charge ordering should lead
to oscillations in the potential drop along the array. Such an observation 
would be a clear evidence of correlated motion. 

\begin{figure}
\includegraphics[width=3.2 in]{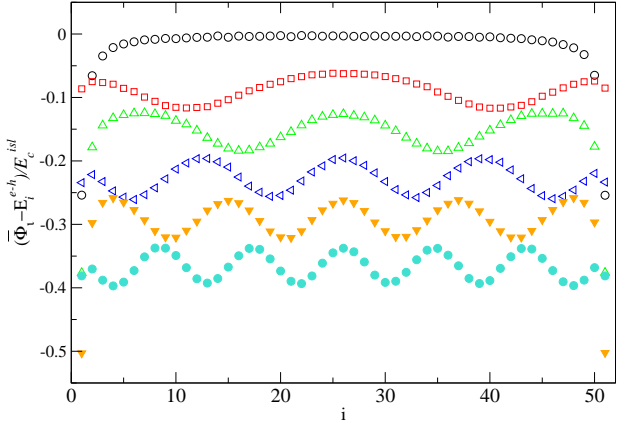}
\caption{Average potential drop, with the excitonic energy substracted,
$\bar \Phi_i-E_i^{e-h}$ at the junctions as a function of position 
at several values of the bias voltage, for which the current is in the
Coulomb staircase regime, corresponding to a clean $N=50$ array and 
$\alpha=1/2$. From top to bottom 
$V=102,104,106,108,110,112 E_c^{isl} $
Curves have been vertically displaced to avoid overlap.
The potential drop show almost regular oscillations
which reflect a stationary (but not static) charge ordering, resembling the
one expected in a Wigner crystal (see discussion in ref.\cite{Stopa01}). 
The  number of maxima and minima does not change within a step and increases
in two (for $\alpha=1/2$) from one steps to the next one. The position of
maxima is slightly adjusted within a step. They start 
appearing close to the electrodes and reflect the entrance of charge from
them.  
}
\end{figure}

  The average potential drop (with the excitonic energy subtracted) through 
the array for several voltages 
corresponding to clean N=50 nanoparticle arrays is shown in Fig.~7.
Clear oscillations are seen. Comparing the values of the bias voltage chosen
with the position of the steps in the corresponding I-V curve in Fig.~7, the
number of maxima/minima in the potential drop do not change in a given step
in the Coulomb staircase. For symmetrically biased arrays, they increase in 
pairs from a step to the next
one. For odd/even number of particles, there is always a minimum/maximum at
the center of the array. The other maxima and minima tend to be as equally 
spaced as possible, but this is not exact. Inconmensurability beween the
period of the oscillations and the lattice can distort  equal spacement.
Also, when new maxima or minima
appear they are closer to the source and drain electrodes and move inwards,
producing a movement of the other maxima and minima, with increasing voltage.
This can be taken as a finite size effect of the Wigner crystal state.  As the
number of charges in the array increase with increasing bias voltage 
the amplitude and period of the oscillations decreases, 
approaching the
high-voltage regime for which $\bar \Phi_i-E^{e-h}_i$  is homogeneous. 

\begin{figure}
\includegraphics[width=3.2 in]{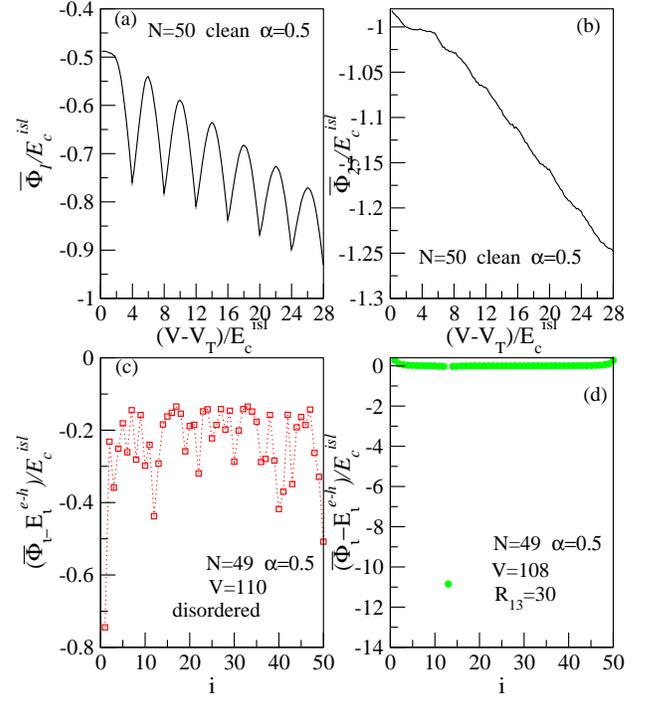}
\caption{(a) and (b) show the average potential drop at the first and 25th
junctions as a function of bias voltage for a clean array with $N=50$ islands, 
$\alpha=1/2$ and no resistance disorder at intermediate voltages. The average 
potential drop at the 
contact junction shows clear oscillations as a function of the bias, while the
potential drop at the center of the array depends monotonously on $V$. 
(c) and (d) show $\Phi_i-E^{e-h}_i$ for charge and resistance disordered 
49-islands arrays respectively and $\alpha=1/2$. The 
Coulomb staircase is much more pronounced in the presence of disorder. 
However, as shown here, the oscillatory
potential drop structure found in Fig.~7 and characteristic of Wigner-crystal 
like physics is destroyed. 
}
\end{figure}

  The anomalous potential drop can be also seen in the potential drop at a 
given junction as a function of the bias voltage, shown in  Fig~8(a) and 8(b) 
for 
junctions 1 and 25 for a clean symmetrically biased 50-islands array  
At the first junction the potential drop show clear oscillations as a function
of the bias voltage, which reflect that new charges state at the first island
are allowed. The potential drop increases until an extra charge can be 
accumulated at the first nanoparticle, for larger voltages the average
occupation of
first island increases and the voltage decreases smoothly until a new value 
at which it increases again as the increase in occupation of first island
cannot compensate the increase in the electrode potential. 
Oscillations, but less regular and less pronounced due to the 
movement of maxima and minima discussed above, are also observed at 
intermediate junctions . Potential drop is much more homogeneous at the middle 
of the array, 
where there is always a minimum (or maximum) in the potential drop.

    Charge or resistance disorder alters the charge motion and frustrates the 
formation of this Wigner crystal like state, as seen in Fig.~8(c) and 8(d). 
This is
the opposite behavior that would be naively expected if one just associates
the appearance of plateaux with the oscillations in voltage drop and emphasizes
 that the step profile is just a consequence of the dependence on the 
bias voltage of the tunneling rate of the processes which control the current.

  The r.m.s. of the junction potential drop is larger than at low voltages and 
smaller than at high voltages. It is of 
the order of the excitonic energy and  reflects the variation in occupation of 
the island. It slowly increases with voltage.

\section{Summary}

In this paper we have studied, analytical and numerically, the current through 
an array of $N$ metallic
islands
placed in between two large electrodes, the source and the drain at 
voltages $V_0$ and $V_{N+1}$ respectively. The applied bias voltage $V$ 
satisfies 
$V=V_0-V_{N+1}$. Interactions are 
restricted to those charges in the same conductor. The capacitive coupling 
between different conductors vanishes. The nanoparticle level spacing
is assumed negligible and transport is treated at the sequential tunneling 
level. In the model used we take into account that the electrodes are not 
ideal voltage sources, but have a finite self-capacitance. This means that the 
voltage at the electrodes fluctuates in response to tunneling processes, but
we assume that prior to the next tunneling event the nominal voltage is 
restored.
Due to the large 
value of the electrode capacitance numerical results are barely modified and 
its effect is neglected in the discussion. The probability of each process 
depends on the change in energy involved, which 
can be written as the energy to create an electron-hole pair plus the 
difference in potential between the sites involved in the tunneling, see 
Eq.(\ref{change}). We have analyzed and clarified how the transport properties 
depend on the number of 
particles, the presence of charge or resistance disorder and the bias voltage, 
including how symmetrically it is applied. To quantify this symmetry we have
introduced a parameter $\alpha$ as $V_0=\alpha V$.

We have shown that in the purely onsite interaction limit the dependence of 
the threshold voltage $V_T$ of clean arrays on the number of nanoparticles 
differs 
qualitatively of the dependence predicted for weakly coupled islands, as in the
onsite case studied here a charge cannot flow freely through an empty array. 
For symmetrically biased arrays $V_T$ equals $2NE_c^{isl}$ for odd $N$ and 
$2E_c(N-1)^{isl}$ for even $N$. The even-odd  effect disappears for forward 
biasing 
($\alpha=1,0$) and $V_T=E_c^{isl}(2N-1)$, see Fig.~1(a). The threshold voltage 
is not affected by 
disorder in the junction resistances but it depends on the selected disorder
configuration if charge disorder is present. With charge disorder, the 
average threshold voltage is independent on $\alpha$ and we recover previously 
predicted values $V_T=E_c^{isl}N$, see Fig.~1(b).

At voltages marginally close to threshold, current is linear on 
$(V-V_T)$ with a
slope independent on the number of particles  $N$ but which depends on the 
resistance of the contact junctions and on the bias asymmetry $\alpha$, see 
Eqs. (\ref{primeq}) to (\ref{ultimaeq}) and Fig.~2. This 
dependence has been obtained both numerical and analytically and resolves 
previous controversy on the power-law close to threshold. 
It reflects that the junction through which charges enter into the array acts
as a bottle neck. 
The range of 
voltages at which this
linear dependence holds is probably too small to be observed experimentally. 
Linearity is lost when the scattering rate for tunneling through the contact 
junctions is comparable to other tunneling processes scattering rates.

The linear regime is followed by a Coulomb staircase at intermediate voltages. 
The width of the steps depends on $\alpha$ and on the presence of charge 
disorder.  For clean arrays
the bias voltage step width is $2E_c^{isl}$ for forward bias and $4E_c^{isl}$ 
for symmetric
bias. 
The stepwidth changes if disorder is present but still depends on the value of 
$\alpha$.
The staircase profile depends on the junction resistances values. See 
Fig.~3.

At high voltages current depends linearly on bias voltage. The asymptotic I-V 
characteristic is given by (\ref{currenthigh}) and cuts the zero current 
axis at a finite 
offset voltage, see Fig.~4. The slope of the asymptotic linear I-V is given by 
the inverse of the sum of the junction resistances in series and the offset 
voltage takes into account charging effects. The high-voltage linear behavior 
is
reached for bias voltages $V_{linear}$ approximately three times larger than 
the offset. $V_{linear}$ can be very large long arrays.

We have studied the potential drop through the array and showed that in none
of the transport regimes studied it drops completely linearly through the 
array. 
In the low-voltage linear regime the average potential drop mainly reflects the
charge gradient created at threshold, see Fig. 6. The effect of disorder in
resistances is extremely weak, except for symmetrically biased arrays with even
number of particles. In the Coulomb staircase regime the potential drop along
the array show almost periodic oscillations in disorder-free arrays which
reflect the correlated motion and the tendency to form a Wigner-crystal like 
state, see Fig.~7. Such periodicity is destroyed by charge or resistance 
disorder in Fig.~8.  
In the high-voltage regime the ohmic dependence, and the associated 
proportionality between the potential drop and the junction resistance ,
is only recovered  once the excitonic energy is substracted, as seen in 
(\ref{potdrophigh}) 
and Fig.~5. The mean value of the potential drop serves to compute the I-V 
characteristic in this regime.

\section{Appendix: Simulation}

We numerically determine the time evolution of the state of an array of 
nanoparticles sandwiched between two large metallic leads symmetrically 
biased at potentials $V_0$ and $V_{N+1}$, with $V_0-V_{N+1}$. The state of 
the array consists of 
the set of charges $\{Q_\beta\}$ that occupy the array islands and the leads. 
The island charges take on integer values. The charge of the nanoparticles is 
modified when an electron tunnels between two adjacent sites. The charges of 
the source and drain can take on any real value because they can be modified 
discretely via the tunneling of charges or continuously via the charging of 
the leads by a battery. 

The evolution of the system is computed by means of a kinetic Monte Carlo 
simulation\cite{Bakhalov89}. 
 At each iteration a single tunneling event takes place. 
The
time involved in this event $\tau$ depends on the 
tunneling rates of all the possible tunneling processes.
Each iteration starts from an initial charge configuration.
First, it is computed the change in energy and 
the tunneling 
rate of the $2(N+1)$ possible hopping events, corresponding to the 
tunneling of a 
single electron, to the left or to the right, 
through any of the $(N+1)$ junctions. The probability of changing the initial 
configuration varies with time like
\begin{equation}
P^{change}(t)=1-P^{stay}(t)=1-e^{-\Gamma^{tot}(t - t_0)}
\label{pmove}
\end{equation}
with $t_0$ the time at which the preceding tunneling process took place and 
$\Gamma^{tot}=\sum_{i=1}^{N+1}(\Gamma^{+}_i+\Gamma^{-}_i)$. $\Gamma^{+}_i$ and 
$\Gamma^{-}_i$ are the tunneling rates through the $i$ junction to the left or 
to the right, 
respectively, 
and 
are calculated from 
(\ref{tunnrate}). To sample the time interval between two hopping events we 
generate random numbers between $[0,1]$ to mimic $P^{change}$ and obtain 
$\tau=t-t_0$ from (\ref{pmove}). 
As the average of $-ln P^{stay}$ is the unity, if one is interested 
only in the average values of the charge or the 
current, and not on 
its fluctuations, the time step could be fixed\cite{Efros03} to 
$1/\Gamma^{tot}$.
This option is numerically faster.

The relative probability of each tunneling event is
$\Gamma^{\pm}_i/\Gamma^{tot}$. To determine the hopping process which 
changes the charge state, 
the relative probabilities are consecutively arranged in the interval 
$[0,1]$.
A second random number in this interval is generated to select the 
tunneling process.  

Then, the charge configuration is updated. After we modify the state of the 
system, we allow the external circuit to return the leads to their applied 
bias values prior to the selection of the next hop. This effect is simulated 
by resetting the charges on the source and the drain to the values that 
restore the nominal applied bias.

In order to remove all sensitivity to initial conditions, before we track the 
evolution of $\{Q_i\}$ as a function of time at any voltage, we perform 
$N_{eq} \ge 10^4$ iterations 
to equilibrate the system.  Following these iterations, we track the evolution 
of the 
charge state until the net number of electrons that arrive at the drain, 
$Q_{drain}$,
equals a very large number ($\ge 10^5$).
The average calculated current is given by
\begin{equation}
I=\frac{Q_{drain}}{t_{tot}}
\label{draincurrent}
\end{equation}
where $t_{tot}$ is the sum of all time intervals between hopping processes in 
the 
evolution runs. 
If a tunneling event involving the drain is selected, an 
amount $\delta q=\pm 1$ is added to $Q_{drain}$ depending on whether an 
electron hopped to or from the drain. Current conservation ensures that the 
average current is the same through any junction. The minimum numbers of
equilibration cycles, $N_{eq}$, and evolution cycles (set by $Q_{drain}$)
depend on the voltage. 

To calculate the average voltage drop, we assume that the system is in a given
state a time equal to the interval $\tau$ until the next tunneling 
event takes place.

Financial support from the Swiss National Foundation, NCCR MaNEP of the Swiss 
National Fonds, the Spanish Science and Education Ministry through Ram\'on y 
Cajal contract and FIS2005-05478-C02-01 grant and the Direcci\'on General de 
Universidades e Investigaci\'on 
de la Consejer\'{\i}a de Educaci\'on de la Comunidad de Madrid and CSIC 
through grant 200550M136  is gratefully acknowledged.
Work at UT Austin was supported by the Welch Foundation by the NSF under grant 
DMR-0606489 and by the ARO under grant W911NF-07-1-0439.


\begin{thebibliography}{99}
\bibitem{cap} Some previous work on capacitance disorder can be found 
in\cite{Melsen97,Jha05}
\bibitem{numeron} In\cite{Middleton93} the average
threshold voltage is predicted to
be linear on the number of junctions, which in our case is $N+1$. It is 
straightforward to argue for both $\alpha=1,1/2$ that $\bar V_T$ is linear in 
the number of bulk islands $N$ as observed in Fig.1(b). In large arrays the
difference between both predictions is negligible.
\end{thebibliography}
\end{document}